\documentclass{nrc1}

\usepackage{cite}
\usepackage{bm}
\newcommand{\vek}[1]{\mathrm{\bf #1}}

\journal{Can. J. Phys.}

\begin{document}

\title{Ultrahigh energy cosmic rays from collisional annihilation revisited}

\author{R. Dick}

\address{Department of Physics and Engineering Physics,
        University of Saskatchewan,\\ 116 Science Place, Saskatoon, 
        SK S7N 5E2, Canada}

\correspond{rainer.dick@usask.ca}

\author{K.M. Hopp}

\author{K.E. Wunderle}

\shortauthor{R. Dick, K.M. Hopp, K.E. Wunderle}

\maketitle

\begin{abstract}
We re-examine collisional annihilation of superheavy dark matter
particles in dark matter density spikes in the galactic halo as a 
possible source of ultrahigh energy cosmic rays. We estimate the 
possible flux in a way that does not depend on detailed assumptions 
about the density profiles of dark matter clumps. The result 
confirms that collisional annihilation is compatible with annihilation 
cross sections below the unitarity bounds for superheavy dark matter 
if the particles can form dense cores in dark matter substructure,
and it provides estimates for core sizes and densities.
The ensuing clumpy source distribution in the galactic halo 
will be tested within a few years of operation of the Pierre 
Auger observatory.

\end{abstract}


\section{Introduction}\label{sec:intro}

The source of the observed ultrahigh energy cosmic rays (UHECRs) with 
energies beyond the Greisen-Zatsepin-Kuzmin cutoff, 
$E>E_{GZK}\simeq 4\times 10^{19}\,$eV, remains a puzzle for 
several reasons:\\
{\bf 1.} Scattering off cosmic microwave
background photons limits the penetration depths of charged
particles at these energies to distances $<100\,$Mpc
\cite{GZK,stecker,new};\\
{\bf 2.} the distribution of arrival directions of UHECRs
does not seem to favor any known astrophysical sources within
the GZK cutoff length;\\
{\bf 3.} it seems extremely difficult to devise sufficiently efficient 
astrophysical acceleration mechanisms which could accelerate particles 
to energies $E>E_{GKZ}$;\\
see e.g. \cite{luis,floyd,tom} for recent reviews.
 For  the second point, we note that Dolag et al. \cite{dolag} find
that typical deflection angles due to magnetic fields should remain below
the angular resolution of the cosmic ray observatories for UHECRs 
originating within $100\,$Mpc, but the pattern of arrival directions does
not seem to point to any known astrophysical accelerators in that range
(but see \cite{sigl} for a different opinion on the possible size
of deflection angles in bottom-up models).
It has been pointed out that there may be a correlation with BL Lac
objects at cosmological distances \cite{tkachev}. However, generation
in BL Lacs would require both an extremely powerful acceleration mechanism
and conversion into a neutral component to avoid the GZK cutoff.
This difficulty is avoided in ``local'' bottom-up scenarios like
e.g. UHECR generation in local gamma-ray bursts within a radius
of $90\,$Mpc \cite{waxman}, but Scully and Stecker had pointed out
that gamma-ray bursts will not provide the observed flux \cite{floyd2}.
Proposals of non-local explanations for absence of a GZK cutoff
include a possible violation of Lorentz invariance at high energies, see
\cite{floyd3} for a recent discussion.
A recent examination of angular correlations by Gorbunov and Troitsky
confirmed the absence of correlations with nearby visible objects \cite{GT}.

The difficulties to explain extremely high energetic cosmic rays had 
motivated Hill already in 1983 to propose monopolonium decay as a possible 
source \cite{Hill}. More recently the constraints 
{\bf 1-3} had also motivated Berezinsky, Kachelrie\ss\ and Vilenkin \cite{BKV} 
as well as Kuzmin and Rubakov \cite{KR} to propose decay of superheavy dark 
matter particles as sources of the observed UHECRs, see also \cite{BS}. The 
terrestrial flux should then be dominated by decays in our galactic dark 
matter halo. Decay of superheavy dark matter is an attractive 
proposal because it explains UHECRs from simple conversion of rest mass 
energy in our cosmic neighborhood, thus evading the GZK bound and the 
need for extremely powerful astrophysical sources. However, it is 
difficult to identify decay mechanisms of particles of mass 
$M_X\ge 10^{12}\,$GeV which are slow enough to ensure a lifetime 
$\tau_X\ge 10^{10}$ years. This had motivated the authors of 
Refs. \cite{BDK1,BDK2,DBK} to propose collisional annihilation of free 
superheavy dark matter particles (SHDM particles, {\sc wimpzillas}) as 
another mechanism to generate UHECRs in our cosmic neighborhood.

The lifetime problem is avoided in the annihilation scenario because the
conversion of rest mass energy into UHECRs proceeds through two-particle
collisions, thus providing a kinematic suppression of the reaction
rate $\dot{n}_X\propto -n_X^2\langle\sigma_A v\rangle$. The
three UHECR puzzles are solved as in the decay 
scenarios: direct conversion of 
rest mass energy eliminates the need for an extremely powerful acceleration 
mechanism; the UHECRs originate within our galactic halo; and the 
distribution of arrival directions is determined by dense dark matter 
subclumps in the galactic halo, rather than by the locations e.g. of 
active galactic nuclei. 

However, the collisional annihilation scenario suffers from its own
specific shortcoming: the minuteness of the annihilation cross section
$\sigma\sim M^{-2}$ of supermassive particles on the one hand ensures 
that these particles can survive to the present day (even with
about equal amounts of matter and anti-matter), but on the
other hand requires high-density substructure in the galactic halo
to generate the observed flux. The results of \cite{BDK1} used two 
separate models for dark matter subhalo profiles and indicated a 
strong dependence of the flux on the shape of the dark matter clumps. 
This left open whether collisional annihilation could be a generic 
possibility or might be constrained by strong restrictions on subhalo 
profiles. We present a simplified estimate of the flux from collisional 
annihilation in SHDM clumps to answer this question. The main purpose of
our simplified estimate is to clarify the dependence of the expected flux
from SHDM annihilation on key parameters like core densities and core 
radii of dark matter subhalos in the galactic halo.
We find that the hypothesis of collisional SHDM annihilation is 
compatible with the observed flux if the cores of dark matter subclumps 
of the galactic halo can reach densities of order $10^{-1}\varrho_\odot$ 
and extensions of a few AU. This corresponds to a mass 
$\sim 10^7M_\odot$ and would indicate that about 10\% of the mass of the 
smallest (and most abundant) substructures detectable with $N$-body 
simulations would be bound in dense cores\footnote{Observational
support for the existence of CDM subhalos within galactic halos
comes from gravitational lensing \cite{chiba,DK}.}.
We denote these dense cores as ``{\sc wimpzilla} stars".

Besides the lifetime constraint for relic decay models and the cross 
section constraint for annihilation, top-down scenarios in general may 
be constrained by indications that the highest energy cosmic rays could
be hadron dominated, see e.g. \cite{watson}. On the other hand, careful 
re-analysis showed that the observed highest energy event could have 
arisen from a primary photon \cite{photon}. 

Photon domination would be expected e.g. if superheavy neutrinos 
annihilate through an intermediate $Z$ or a Higgs boson, because the 
ensuing $\pi^0$ shower would decay into UHE photons over mean free paths 
of order $10\,$km. The composition of showers in such a model would
therefore resemble $Z$-bursts, which assume annihilation of ultrahigh 
energy neutrinos (instead of supermassive neutrinos) through scattering 
with background neutrinos \cite{zburst,ringwald,tomw}.
However, the different origin of $Z$-bursts implies a markedly different
anisotropy signature: $Z$-bursts at relic neutrinos in our GZK volume
would produce an isotropic pattern of arrival directions without
connection to the galactic halo.

The first crucial experimental test for our top-down model
of SHDM annihilation in {\sc wimpzilla} stars will be the 
observed anisotropy signature after a few years of operation of the 
Pierre Auger observatory \cite{kampert}:\\
If our annihilation model is correct, a pointlike source distribution
should emerge with increasing source density towards the galactic
center \cite{DBK}. In particular, the majority of observed UHE events
after about four years should appear in doublets or higher multiplets.\\
Alternatively, if the decay model is correct, the anisotropy 
signature after about five years should trace the galactic dark matter 
halo with uniformly increasing intensity towards the galactic center.

In Sec. \ref{sec:gen} we discuss the generation of weakly coupled 
particles during inflation and explain why stable particles with masses
$m\ge 10^{12}\,$GeV are still around to create the observed UHE cosmic
rays through annihilation.
In Secs. \ref{sec:basic} and \ref{sec:flux} we simplify the estimates
from \cite{BDK1} for the UHE cosmic ray flux which might be expected 
from SHDM annihilation in dark matter substructures within our galactic 
halo. The method developed in Secs. \ref{sec:basic} and \ref{sec:flux} 
yields improved estimates of typical core sizes and densities
of these substructures. Sec. \ref{sec:basic} reviews the basic formalism 
for the calculation of cosmic ray fluxes from dark matter annihilation,
and in Sec. \ref{sec:flux} we present our simplified estimate of the 
flux from SHDM annihilation. 
We set $\hbar=c=1$ except in equations which are used for 
numerical evaluations. The differential flux per energy interval
is $j(E)=dI/dE$, and the flux per unit of solid angle is
$J(E)=d^2I/dEd\Omega$.

 \section{Superheavy dark matter from non-adiabatic expansion}\label{sec:gen} 

One of the most interesting sources for superheavy dark matter particles
is gravitational production during non-adiabatic expansion. This 
mechanism is usually considered in terms of the Bogolubov transformation
between in and out vacua \cite{parker,davies}, and was also extensively 
discussed as a source of superheavy dark matter particles 
\cite{BCKL,MR,KRT,CKR,KT,GPRT,CKRT,CCKR}. Here we would like to point out 
how this mechanism for particle production can be understood in terms of 
the evolution equation for weakly coupled helicity states in an expanding
universe. In the present paper we will discuss this only for weakly coupled
helicity states which are not ultra-relativistic ($\lambda\not\ll m^{-1}$),
although we are sure that similar conclusions can be drawn for the
ultra-relativistic case, too.

The direct coupling between non-adiabacity and changing particle numbers
during gravitational expansion follows directly from the Einstein 
equation and the cosmological principle, since it is well known that 
the evolution equations for the Robertson--Walker metric
\begin{equation}\label{eq:friedmann}
\frac{\dot{a}^2+k}{a^2}=\frac{\kappa}{3}\varrho
\end{equation}
and
\[
2\frac{\ddot{a}}{a}+\frac{\dot{a}^2+k}{a^2}=-\kappa p
\]
imply energy balance with only mechanical energy:
\begin{equation}\label{eq:energy1}
d(\varrho a^3)=-pda^3,
\end{equation}
and therefore the first law of thermodynamics splits into 
Eq. (\ref{eq:energy1}) and a separate balance equation for entropy 
and particle densities:
\begin{equation}\label{eq:entropy1}
Td(s a^3)=-\sum_i\mu_i d(\nu_i a^3).
\end{equation}

At the same time Eq. (\ref{eq:friedmann}) usually implies a decrease
or at best constancy of the comoving energy density $a^3(t)\varrho(t)$ 
during expansion. Consider e.g. the standard linear dispersion relation 
for radiation or dust 
\begin{equation}\label{eq:disprel1}
p=\left(\frac{\ell}{3}-1\right)\varrho,
\end{equation}
where the parameter is $\ell=4$ for radiation and $\ell=3$ for dust.
After inflation, when $k$ can be neglected and radiation initially
dominates, the scale factor and dominant energy density evolve
according to $a(t)\propto t^{1/2}$ and $\varrho(t)\propto t^{-2}$,
leading to a decrease 
$a^3(t)\varrho(t)\propto t^{-1/2}$ of the comoving energy density.
However, during inflation we have $p\simeq -\varrho$ leading to
constant energy density $\varrho$ and an exponential increase
$a^3(t)\varrho(t)\propto\exp(3Ht)$ of the comoving energy density.
 From this argument we expect an energy increase at least in all
those states which cannot readily redistribute their energy through
sufficiently strong interactions, and
inspection of the corresponding evolution
equations for weakly coupled helicity states confirms this:\\
Towards the end of inflation the curvature parameter $k$ is negigible
on subhorizon scales, and weakly coupled helicity states have to satisfy
\begin{equation}\label{eq:phi1}
\ddot{\phi}(\vek{k},t)+3H\dot{\phi}(\vek{k},t)
+\left(m^2+\vek{k}^2\exp(-2Ht)\right)\phi(\vek{k},t)=0.
\end{equation}
Here $\phi(\vek{k},t)$ is defined through Fourier
expansion with respect to the dimensionless comoving
coordinates $\vek{x}$
\[
\phi(\vek{x},t)=\frac{1}{\sqrt{2\pi}^3}\int\!d^3\vek{k}\,
\phi(\vek{k},t)\exp(\mathrm{i}a_1\vek{k}\cdot\vek{x}).
\]
The factor $a_1=a(t_1)\exp(-Ht_1)$ is related to the scale factor $a(t_1)$ 
at the beginning of inflation.
Towards the end of inflation the exponential factor in Eq. (\ref{eq:phi1})
is exponentially small, $\exp(-2Ht)\approx\exp(-140)$, and we find the
same approximate evolution equation for all modes which are not 
ultrarelativistic ($|\vek{k}|\not\gg m$) at the end of inflation:
\begin{equation}\label{eq:phi2}
\ddot{\phi}(\vek{k},t)+3H\dot{\phi}(\vek{k},t)
+m^2\phi(\vek{k},t)\simeq 0.
\end{equation}
$H$ is constant during inflation, and therefore the time evolution of 
the comoving energy density in the weakly coupled helicity states is given by
\begin{eqnarray*}
a^3(t)\varrho_{\phi}(\vek{k},t)&\simeq&
\frac{1}{2}a^3(t)\left(\dot{\phi}(\vek{k},t)\dot{\phi}(-\vek{k},t)
+m^2\phi(\vek{k},t)\phi(-\vek{k},t)\right)
\\
&\simeq&
A_+\exp\!\left(t\sqrt{9H^2
-4m^2}\right)+A_-\exp\!\left(-t\sqrt{9H^2
-4m^2}\right)+B.
\end{eqnarray*}
This implies a growing mode in the comoving energy density of weakly coupled
states with $m<1.5H\simeq 10^{14}\,$GeV. 
What is special about the superheavy particles is that their 
comoving energy density is conserved after inflation, because the 
behavior of massive $(m>t^{-1})$ weakly coupled helicity states 
in the subsequent radiation and dust dominated backgrounds preserves 
their energy:\\
The asymptotic solution for weakly coupled massive states 
with $m>t^{-1}$ in such a background yields
\[
\phi(\vek{k},t)\propto
t^{-3/\ell}\cos\!\left(mt+\varphi\right),
\]
\[
\varrho_{\phi}(\vek{k},t)\propto t^{-6/\ell}\propto a^{-3},
\]
and this implies in particular that the comoving density of massive 
particles freezes out
at the end of inflation $(t\simeq 10^{-36}\,\mbox{s})$ if 
\[
m>t^{-1}\simeq 10^{12}\,\mbox{GeV}.
\]
 From these considerations follows a mass window for inflationary production
of superheavy relic particles 
\[
10^{12}\,\mbox{GeV}<m<10^{14}\,\mbox{GeV}.
\]
It is certainly reassuring for the top-down models that a typical time 
scale for the duration of inflation yields the UHECR mass scale for the
preserved particles.

\section{Basic formalism}\label{sec:basic}

 Following \cite{BDK1}, we consider annihilation of two superheavy
dark matter particles of mass $M_X\ge 10^{12}\,$GeV 
primarily into two jets of energy $M_X$.

The cosmic ray flux per energy interval originating from 
annihilation of particles and anti-particles of total density 
$n_X(\vek{r})$ is 
\begin{equation}\label{eq:I}
j(E,\vek{r}_{\odot})=\frac{dI(E,\vek{r}_{\odot})}{dE}
=\frac{d\mathcal{N}(E,M_X)}{dE}\int d^3\vek{r}\,
\frac{n_X^2(\vek{r})\langle\sigma_A v\rangle}{8\pi
|\vek{r}_{\odot}-\vek{r}|^2}
\end{equation}
if the particles are {\it not} Majorana particles, or 
$4\times j(E,\vek{r}_{\odot})$ otherwise.
$d\mathcal{N}(E,M_X)$ is the number of particles
in the energy interval $[E,E+dE]$ emerging from a jet of energy 
$M_X$ and two primary jets are assumed. Neglecting events with more 
than two jets, $d\mathcal{N}(E,M_X)/dE$ is related to fragmentation 
functions via
\[
\frac{d\mathcal{N}(E,M_X)}{dE}=\sum_i\frac{1}{2\sigma_A}
\frac{d\sigma^{(i)}}{dE}
=\frac{1}{2M_X}\sum_i F^{(i)}(x,4M_X^2),
\]
where $x=E/M_X$ and $F^{(i)}(x,4M_X^2)$ is the differential
number of particles of species $i$ generated in the prescribed 
$x$-range in an annihilation event with $s=4M_X^2$.

As shown in \cite{BDK1}, UHECRs from annihilation of superheavy dark 
matter particles provide already a good fit to the shape of the observed 
spectrum if the old MLLA fragmentation function 
\cite{mlla} is used for an estimate on $d\mathcal{N}(E,M_X)/dE$.
More refined approximations have been employed
in the discussion of the decay scenario \cite{BS,FK,ST,BD,ABK}, and these also 
further improve the corresponding fit for the annihilation scenario, 
since for the spectral shape the transition between the decay and 
annihilation scenarios only corresponds to a scaling of jet energies
\begin{equation}\label{eq:jetscale}
\left.\frac{d\mathcal{N}(E,M_X/2)}{dE}\right|_{decay}
\to\left.\frac{d\mathcal{N}(E,M_X)}{dE}\right|_{annihilation}
\end{equation}
in the fragmentation function. 

\section{A simplified estimate of the flux from dark matter clumps 
in the galactic halo}
\label{sec:flux}

Numerical simulations indicate that the dark matter
halos of galaxies have a lot of substructure in terms
of spatial density fluctuations. In \cite{BDK1} it was
found that the flux from local overdensities, or
dark matter clumps, exceeds the flux from the smooth
halo component by far. Here we will reconsider these
calculations with a view on the density $n_X$
and size $r_{core}$ of the central core regions of dark matter clumps. 
$n_X$ is of crucial importance for the UHECR flux from annihilation 
in these clumps.

We also want to simplify the flux estimate with regard to assumptions
on the distribution of subclumps. In \cite{BDK1} a clump distribution
\begin{equation}\label{eq:assum1}
n_{cl}(d,M_{cl})\propto\left(\frac{M_{cl}}{M_H}\right)^{-\alpha}
\left[1+
\left(\frac{d}{d_0}\right)^2\right]^{-3/2}
\end{equation}
was assumed, where $n_{cl}(d,M_{cl})$ is the volume density of clumps
with mass $M_{cl}$ at distance $d$ from the galactic center,
$d_0$ is a scale radius for the subclump distribution in the galaxy,
and $M_H$ is the mass of the galactic halo \cite{blasisheth}.
Some of the recent work on structure formation has questioned the assumption
of increasing clump density towards the galactic center because tidal
stripping would more severely affect large clumps close to the galactic
center. However, the distribution of the 150 known globular clusters
in the galaxy suggests an increase of visible substruture towards the 
galactic center \cite{harris},
and it seems reasonable to assume a similar distribution
for dark matter substructure.

In the present paper, we want to avoid any detailed assumptions 
like (\ref{eq:assum1}) about the profile of the substructure distribution
in the galactic halo, and we also want to avoid any
assumptions about the density profiles of substructure, except for the
reasonable assumption that the substructures will have dense
cores with an average mass density $n_XM_X$.

The annihilation flux from an approximately compact source 
of volume $V$ at distance $d(\gg V^{1/3})$ is
\begin{equation}\label{eq:singlesource}
j(E,\vek{r}_{\odot})
=\int_V d^3\vek{r}\,\frac{1}{8\pi|\vek{r}_{\odot}-\vek{r}|^2}
\frac{d\mathcal{N}(E,M_X)}{dE}
n_X(\vek{r})^2 \langle\sigma_A v\rangle 
\simeq\frac{V}{8\pi d^2}\frac{d\mathcal{N}(E,M_X)}{dE}
n_X^2 \langle\sigma_A v\rangle.
\end{equation}

UHECR experiments often plot the approximately flat
rescaled flux 
\[
E^3j(E,\vek{r}_{\odot})
\simeq\frac{E^3 V}{8\pi d^2 M_Xc^2}
\frac{d\mathcal{N}(x,M_X)}{dx}
n_X^2 \langle\sigma_A v\rangle,
\]
where $x=E/M_Xc^2$ is energy in units of primary jet energy.
A typical value for fragmentation functions at $x=0.1$ is 
$d\mathcal{N}(x,E_{\mathrm{jet}})/dx\simeq 30$.

In the following we parametrize the annihilation cross section in terms
of the s-wave unitarity bound for $Mc^2=10^{12}\,$GeV and 
$v=100\,$km/s:
\begin{equation}\label{eq:sigma}
\langle\sigma_A v\rangle=\xi\times\frac{4\pi\hbar^2}{M^2v}
=\xi\times 4.40\times 10^{-43}\,\mbox{m}^3/\mbox{s},
\end{equation}
where the unitarity bound is $\xi\le 1$. 

For a fiducial average distance we use $d=10\,\mbox{kpc}$, 
whereas for the fiducial volume and density of subclump cores
we should have
\[
n_X M_X \overline{V}_{core}\simeq
0.1\overline{M}_{cl}=0.1f_{cl}M_{halo}/N_{cl}
\simeq 10^{-5}M_{halo}. 
\]
Here $f_{cl}$ is the mass fraction in the $N_{cl}$ dark matter subclumps.
Numerical $N$-body simulations and gravitational lensing indicate
that $f_{cl}$ should be of the order of a few percent.
E.g. a core density
\begin{eqnarray*}
n_X M_X&=&\eta\varrho_\odot=\eta\times 1.41\times 10^3\,
\mbox{kg}/\mbox{m}^3,
\\
n_X&=&\eta\times\frac{\varrho_\odot}{10^{12}\,\mbox{GeV}/c^2}
=\eta\times 7.89\times 10^{17}\,\mbox{m}^{-3},
\end{eqnarray*}
yields typical core volumes and radii
\begin{eqnarray*}
\overline{V}_{core}&\simeq&
\eta^{-1}\times 2\times 10^{7}V_\odot,
\\ \nonumber
\overline{r}_{core}
&\simeq&\eta^{-1/3}\times 1.9\times 10^{11}
\,\mbox{m}.
\end{eqnarray*}
With these fiducial average values, the flux at
$E=10^{11}\,$GeV from $N_{cl}$
dense cores of clumps can be estimated as
\begin{eqnarray}\label{eq:flux}
E^3 j(E)\Big|_{E=10^{11}\,\mathrm{GeV}}
&\simeq&
N_{cl}\overline{V}_{core}\frac{E^3}{8\pi d^2 M_Xc^2}
\frac{d\mathcal{N}(x,M_X)}{dx}
n_X^2 \langle\sigma_A v\rangle
\\ \nonumber
&\simeq& 0.1f_{cl}M_{halo}\frac{E^3}{8\pi d^2 M_X^2c^2}
\frac{d\mathcal{N}(x,M_X)}{dx}
n_X \langle\sigma_A v\rangle
\\ \nonumber
&\simeq& 5.83\times 10^{25}\, 
\mbox{eV}^2\mbox{m}^{-2}\mbox{s}^{-1}
\times\frac{f_{cl}}{0.06}
\times\frac{M_{halo}}{2\times 10^{12}M_\odot}
\times\left(\frac{d}{10\,\mathrm{kpc}}\right)^{-2}
\times\frac{\eta\xi}{10^{-3}},
\\
E^3J(E)\Big|_{E=10^{11}\,\mathrm{GeV}}
&\simeq& 4.64\times 10^{24}\,
\mbox{eV}^2\mbox{m}^{-2}\mbox{s}^{-1}\mbox{sr}^{-1}
\\ \nonumber
&&\times\frac{f_{cl}}{0.06}
\times\frac{M_{halo}}{2\times 10^{12}M_\odot}
\times\left(\frac{d}{10\,\mathrm{kpc}}\right)^{-2}
\times\frac{\eta\xi}{10^{-3}}.
\end{eqnarray}
This is compatible with the UHECR flux observed by 
AGASA\footnote{The HiRes collaboration uses a different binning 
of the AGASA data, but still finds a similar AGASA flux at and 
above $E=10^{11}\,$GeV without a cutoff, see e.g. \cite{hires}. 
However, the HiRes collaboration reports a cutoff in the spectrum 
based on its own data. This issue will be resolved by the Pierre 
Auger Observatory due to the much improved statistics expected 
from that experiment, and due to the combination of
Cherenkov detectors and air fluorescence telescopes.}
\cite{agasa}, 
\[
E^3J(E)|_{E=10^{11}\,\mathrm{GeV}}\\
\simeq 4.2
\times 10^{24}\,\mbox{eV}^2\mbox{m}^{-2}\mbox{s}^{-1}\mbox{sr}^{-1},
\]
if the product of the parameters $\eta$ and $\xi$ is of order
$\eta\xi\sim 10^{-3}$. E.g. reproduction of the beautiful 
standard model fragmentation fit in Fig. 7 from Ref. \cite{ST}
with the present approximations would require
$\eta\xi\simeq 8.6\times 10^{-4}$. 

\section{Conclusion}\label{sec:conc} 

The result (\ref{eq:flux}) indicates that collisional annihilation
of superheavy dark matter particles remains a viable scenario for 
the origin of UHECRs, albeit at the expense of postulating dense
cores in dark matter substructure. This difficulty compares
e.g. to the lifetime problem in decay scenarios, or to the problem
of defeating synchrotron radiation loss and collisional energy loss
in bottom-up scenarios.

The unitarity bound (\ref{eq:sigma}) on the annihilation cross
section is both a virtue and a curse of the collisional annihilation
scenario. It is a virtue because it implies that ultrahigh energy
cosmic rays from collisional annihilation must originate in dense
cores of dark matter clumps in the galactic halo, amounting
to a unique prediction: an unmistakable clumpy pattern
in arrival directions should emerge after four to five years of 
operation of the Pierre Auger observatory, appearing as pointlike 
sources within our halo. The connection to the halo should become
apparent through an increase of source density towards the galactic
center, and it will eliminate extragalactic scenarios for UHE origin.
In addition, the clustering of arrival directions distinguishes it
from the SHDM decay scenarios, and the absence of correlations of
the pointlike sources with supernovae will rule out a bottom-up
origin.
A further advantage of the unitarity bound is that it ensures
that stable superheavy dark matter particles will still be around due
to the extremely slow rates of any reactions involving these particles
and whatever affected the low mass modes to generate the baryon asymmetry
after inflation very likely could not affect the superheavy modes.
Yet at the same time the unitarity bound seems to require
high core densities, and this may collisional annihilation look 
less appealing in the current stage, where it is not explicitly ruled 
in or ruled out. However, the required core densities for dark matter
subclumps are not unreasonable and the possibility of UHECRs from
collisional SHDM annihilation can not easily be dismissed.

 From our point of view, primary composition is a secondary criterion
for separation of different scenarios for UHECR origin.
Inferring primary composition from observations requires particle
physics modelling with extrapolations to very high energies
whereas the interpretation of anisotropy will involve much less 
theoretical uncertainty. 

Either way, the anisotropy signature observed by the Pierre Auger
observatory will decide the fate of the annihilation scenario
and we are cautiously optimistic that the collisional SHDM
annihilation mechanism in {\sc wimpzilla} stars will withstand the 
tests of the near future.

\section*{Acknowledgement}

This research was supported in part by the Natural Sciences and Engineering
Research Council of Canada. We are grateful to Floyd Stecker for helpful 
comments on an early draft of this paper.


\begin{thebibliography}{99}
\bibitem{GZK}K. Greisen, Phys. Rev. Lett. {\bf 16}, 748 (1966); 
G.T. Zatsepin, V.A. Kuzmin, JETP Lett. {\bf 4}, 78 (1966).

\bibitem{stecker}F.W. Stecker, Phys. Rev. Lett. {\bf 21}, 1016 (1968).

\bibitem{new}T. Stanev, R. Engel, A. M\"ucke, R.J. Protheroe, 
J.P. Rachen, Phys. Rev. {\bf D62}, 093005 (2000).

\bibitem{luis}D.F. Torres, L.A. Anchordoqui, Rep. Prog. Phys.
{\bf 67}, 1663 (2004).

\bibitem{floyd}F.W. Stecker, J. Phys. {\bf G29}, R47 (2003).

\bibitem{tom}T.K. Gaisser, Nucl. Phys. B (Proc. Suppl.) {\bf 117}, 
 318 (2003).

\bibitem{dolag}K. Dolag, D. Grasso, V. Springel, I. Tkachev,
JETP Lett. {\bf 79}, 583 (2004); JCAP {\bf 0501}, 009 (2005).

\bibitem{sigl}G. Sigl, F. Miniati, T. Ensslin, Phys. Rev. {\bf D70},
 043007 (2004), Nucl. Phys. B (Proc. Suppl.) {\bf 136}, 224 (2004).

\bibitem{tkachev}P.G. Tinyakov, I.I. Tkachev, JETP Lett. {\bf 74},
 445 (2001).

\bibitem{waxman}E. Waxman, Phys. Rev. Lett. {\bf 75}, 386 (1995),
{\it Gamma-ray bursts: Potential sources of ultra high energy cosmic
rays}, astro-ph/0412554.

\bibitem{floyd2}S.T. Scully, F.W. Stecker, Astropart. Phys. {\bf 16}, 
271 (2002).

\bibitem{floyd3}F.W. Stecker, S.T. Scully, Astropart. Phys. {\bf 23},
203 (2005).

\bibitem{GT}D.S. Gorbunov, S.V. Troitsky, Astropart. Phys. {\bf 23},
175 (2005).

\bibitem{Hill}C.T. Hill, Nucl. Phys. {\bf B224}, 469 (1983).

\bibitem{BKV}V. Bere\-zins\-ky, M. Kachel\-rie\ss, A. Vilen\-kin,
 Phys. Rev. Lett. {\bf 79}, 4302 (1997).

\bibitem{KR}V.A. Kuzmin, V.A. Rubakov, in 
{\it Beyond the Desert --- Accelerator and Non-Accelerator Approaches},
edited by H.V. Klapdor-Kleingrothaus and H. P\"as, IoP Publishing, 
Bristol 1998, pp. 802--807; Phys. Atom. Nucl. {\bf 61}, 1028 (1998).

\bibitem{BS}
M. Birkel, S. Sarkar, Astropart. Phys. {\bf 9}, 297 (1998).

\bibitem{BDK1}
P. Blasi, R. Dick, E.W. Kolb, Astropart. Phys. {\bf 18}, 57 (2002).
\bibitem{BDK2}
P. Blasi, R. Dick, E.W. Kolb, Nucl. Phys. B (Proc. Suppl.) {\bf 110}, 
494 (2002).

\bibitem{DBK}R. Dick, P. Blasi, E.W. Kolb, 
Nucl. Phys. B (Proc. Suppl.) {\bf 124}, 201 (2003).

\bibitem{chiba}
M. Chiba, Astrophys. J. {\bf 565}, 17 (2002).

\bibitem{DK}
N. Dalal, C.S. Kochanek, Astrophys. J. {\bf 572}, 25 (2002).

\bibitem{watson}A.A. Watson, {\it The mass composition of
 cosmic rays above $10^{17}\,$eV}, astro-ph/0410514.

\bibitem{photon}
P. Homola et al., Acta Phys. Pol. {\bf B35}, 1893 (2004);
M. Risse et al., {\it On the primary particle type of the most
 energetic Fly's Eye experiment}, astro-ph/0410739.

\bibitem{zburst}
T. Weiler, Astropart. Phys. {\bf 11}, 303 (1999);
D. Fargion, B. Mele, A. Salis, Astrophys. J. {\bf 517},
725 (1999).

\bibitem{ringwald}Z. Fodor, S. Katz, A. Ringwald, 
JHEP {\bf 0206}, 046 (2002).

\bibitem{tomw}G. Gelmini, G. Varieschi, T. Weiler, 
Phys. Rev. {\bf D70}, 113005 (2004).

\bibitem{kampert}K.-H. Kampert (for the Pierre Auger Collaboration),
{\it The Pierre Auger Observatory - Status and Prospects},
astro-ph/0501074.

\bibitem{parker}L. Parker, Phys. Rev. {\bf 183}, 1057 (1969).

\bibitem{davies}N.D. Birrell, P.C.W. Davies, 
{\it Quantum Fields in Curved Space},
Cambridge University Press, Cambridge 1982.

\bibitem{BCKL}J.D. Barrow, E.J. Copeland, E.W. Kolb, 
A.R. Liddle, Phys. Rev. {\bf D43}, 977 (1991).

\bibitem{MR}A. Masiero, A. Riotto, Phys. Lett. {\bf B289}, 73 (1992).

\bibitem{KRT}E.W. Kolb, A. Riotto, I.I. Tkachev, 
Phys. Lett. {\bf B423}, 348 (1998).

\bibitem{CKR}D.J.H. Chung, E.W. Kolb, A. Riotto,  Phys. Rev. {\bf D59},
 023501 (1999), Phys. Rev. {\bf D60}, 063504 (1999).

\bibitem{KT}V.A. Kuzmin, I.I. Tkachev,
 Phys. Rev. {\bf D59}, 123006 (1999), Phys. Rep. {\bf 320}, 199 (1999).

\bibitem{GPRT}
G.F. Giudice, M. Peloso, A. Riotto, I.I. Tkachev, JHEP {\bf 9908}, 014 (1999).

\bibitem{CKRT}D.J.H. Chung, E.W. Kolb, A. Riotto, I.I. Tkachev,
 Phys. Rev. {\bf D62}, 043508 (2000).

\bibitem{CCKR}D.J.H. Chung, P. Crotty, E.W. Kolb, A. Riotto, 
Phys. Rev. {\bf D64}, 043503 (2001).

\bibitem{mlla}Ya.I. Azimov, Yu.L. Dokshitzer, V.A. Khoze, S.I. Troyan, 
 Phys. Lett. {\bf B165}, 147 (1985), Z. Phys. {\bf C27}, 65 (1985),
 Z. Phys. {\bf C31}, 213 (1986).

\bibitem{FK}Z. Fodor, S.D. Katz, Phys. Rev. Lett. {\bf 86}, 3224 (2001).

\bibitem{ST}S. Sarkar, R. Toldr\`{a}, Nucl. Phys. {\bf B621}, 495 (2002).

\bibitem{BD}C. Barbot, M. Drees, Astropart. Phys. {\bf 20}, 5 (2003).

\bibitem{ABK}R. Aloisio, V. Berezinsky, M. Kachelrie\ss,
 Phys. Rev. {\bf D69}, 094023 (2004).

\bibitem{blasisheth}P. Blasi, R. Sheth,
Phys. Lett. {\bf B486}, 233 (2000).

\bibitem{harris}W.E. Harris, Astron. J. {\bf 112}, 1487 (1996).

\bibitem{agasa}M. Takeda et al.,
Astropart. Phys. {\bf 19}, 447 (2003).

\bibitem{hires}R.U. Abbasi et al., Phys. Rev. Lett. {\bf 92},
151101 (2004).

\end{thebibliography}
\end{document}